\documentclass[
twocolumn,
aps,
prl,
reprint,
superscriptaddress,
amsmath,amssymb,
longbibliography
]{revtex4-1}
\usepackage[pdftex]{graphicx}
\usepackage[pdftex,
            colorlinks=true,
            citecolor=blue,
            linkcolor=blue]{hyperref}
\graphicspath{{./fig},{./fig_SM}}

\usepackage{braket}
\usepackage{times,multirow,amsfonts,bm,xspace,pifont}
\usepackage{soul}
\usepackage[normalem]{ulem}
\usepackage{textcomp}

\begin{document}
\newcommand{\sgn}{\mathrm{sgn}\,}
\newcommand{\tr}{\mathrm{tr}\,}
\newcommand{\Tr}{\mathrm{Tr}\,}

\renewcommand{\Re}{\mathrm{Re}}
\renewcommand{\Im}{\mathrm{Im}}

\newcommand{\halpha}{{\hat{\alpha}}}
\newcommand{\uint}{{\int_0^\infty}}
\newcommand{\mA}{{\mathcal{A}}}
\newcommand{\mP}{{\mathcal{P}}}
\newcommand{\mB}{{\mathcal{B}}}
\newcommand{\mJ}{{\mathcal{J}}}
\newcommand{\bx}{{\bar{x}}}
\newcommand{\by}{{\bar{y}}}
\newcommand{\bz}{{\bar{z}}}
\newcommand{\bepsilon}{{\bar{\epsilon}}}

\newcommand{\SC}{{\rm{s}}}
\newcommand{\normal}{{\rm{n}}}

\newcommand{\Tc}{{T_{\rm c}}}
\newcommand{\qc}{{q_{\rm c}}}
\newcommand{\Tcn}{{T_{\rm c0}}}
\newcommand{\RK}{{R_{\rm K}}}

\title{Nonreciprocal Current-Induced Zero-Resistance State in Valley-Polarized Superconductors}

\author{Akito Daido} 
\email[]{daido@scphys.kyoto-u.ac.jp}
\affiliation{Department of Physics, Graduate School of Science, Kyoto University, Kyoto 606-8502, Japan}
\affiliation{Department of Physics, Hong Kong University of Science and Technology, Clear Water Bay, Hong Kong, China}
\author{Youichi Yanase}
\affiliation{Department of Physics, Graduate School of Science, Kyoto University, Kyoto 606-8502, Japan}
\author{K. T. Law}
\affiliation{Department of Physics, Hong Kong University of Science and Technology, Clear Water Bay, Hong Kong, China}
\date{\today}

\begin{abstract}
The recently observed nonreciprocal current-induced zero-resistance state (CIZRS) in twisted trilayer graphene/WSe$_2$ heterostructure has posed a significant theoretical challenge. {In the experiment}, the system shows a zero-resistance state only when a sufficiently large current is applied in a particular direction, while stays in an incipient superconducting state with small resistance when the current is small or {flows} in the opposite direction.
In this Letter, we provide a theory of {CIZRS}. We show that the threefold degenerate Fulde-Ferrell (FF) states are stabilized {by the valley polarization and trigonal warping effects of} twisted trilayer graphene/WSe$_2$ heterostructures. Moreover, a current flowing in a particular direction breaks the threefold degeneracy   and favors a particular FF pairing domain.  We therefore propose that the incipient superconducting state is naturally understood as a multidomain state where the interdomain supercurrent is difficult to flow due to the tiny Josephson coupling caused by the mismatch of Cooper-pair momenta between different FF domains. 
Nevertheless, a sufficiently large current in a particular direction can selectively populate a certain FF state and create monodomain pathways with zero resistance. Crucially, due to the threefold symmetry of the system, a current flowing in the opposite direction {can fail} to generate the zero-resistance pathways, thus giving rise to the observed nonreciprocity. Finally, we suggest that the long-sought-after triangular finite-momentum state can also be realized in valley-polarized superconductors.
\end{abstract}

\maketitle

\textit{Introduction}---Controlling quantum states by electric currents is one of the most intriguing issues in condensed matter physics. Among various systems, superconductors offer a fascinating platform owing to their unique electromagnetic properties. It is widely believed that superconductors show zero resistance {until it is destroyed by a large current}. However, contrary to this paradigm, a recent experiment on twisted trilayer graphene/WSe$_2$ heterostructures reported a surprising situation, {where the sample exhibits zero resistance under large currents but shows a finite resistance, though smaller than its normal-state value, under small currents ~\cite{Lin2022-cz}. This behavior is observed for the current flowing in a particular direction, while the system always exhibits a finite resistance in the opposite direction} [see Fig.~\ref{fig:pd_j_random}(d) for example]. {The observation of such} nonreciprocal current-induced zero-resistance state (CIZRS) unambiguously illustrates the necessity to {deepen} our understanding of superconductors in electric current.

The nonreciprocal behavior of the CIZRS is an analog of the nonreciprocal transport phenomenon called the superconducting diode effect (SDE)~\cite{Ando2020-om,Nagaosa2024-ut}, which was also observed in twisted trilayer graphene/WSe$_2$ with other moir\'e fillings and displacement fields. 
From symmetry viewpoints, {the} SDE requires broken inversion and time-reversal symmetries~\cite{Nagaosa2024-ut},
whose origin in twisted trilayer graphene/WSe$_2$ is proposed to be a spontaneous occupation imbalance of the electrons' valley degree of freedom~\cite{Zhang2024-ec, Lin2022-cz, Siriviboon2021-un,Scammell2022-pv}, i.e., the valley polarization~\cite{Ajayan2016-cm, Novoselov2016-ev, Geim2013-vc, Novoselov2005-nv, Novoselov2004-pn, Christos2022-ax, Bultinck2020-yc, Zhang2019-xe,Hu2025-xc}. It is natural to consider that the valley polarization plays a key role in {the} CIZRS as well.

There have been several attempts to explain the CIZRS. As commented on in Refs.~\cite{Scammell2022-pv,Zhuang2025-ks}, CIZRS could be realized if the valley polarization was sufficiently weakened by applying electric current, because it is generally disadvantageous for superconductivity. To demonstrate this idea, superconducting instability  must be studied in dissipative electric current. Recent effort along this line, however, has not succeeded to obtain CIZRS~\cite{Banerjee2025-xi}. {Another possible strategy to obtain CIZRS is to drive the system out of equilibrium~\cite{Daido2025-si}}, but its connection to twisted trilayer graphene/WSe$_2$ {remains} unclear. Thus, there is no satisfactory theory explaining CIZRS in twisted trilayer graphene/WSe$_2$.

In this Letter, we argue that not only CIZRS but also its nonreciprocal behavior can be a natural property of superconductors with valley polarization and trigonal warping of Fermi surfaces.
With a minimal model,
we show that trigonal warping stabilizes  the Fulde-Ferrell (FF) states, i.e., finite-momentum superconductivity with a plane-wave order parameter~\cite{Fulde1964-qq}.
The obtained FF states show threefold degeneracy, and can be externally switched by taking advantage of the coupling between Cooper-pair momentum and electric current.
By establishing the electric-current phase diagram of FF states, we propose a simple scenario of CIZRS by considering FF domains and weak Josephson couplings between them due to the mismatch of Cooper-pair momenta. In particular, our scenario explains three key aspects of the experiment: (1) the incipient superconducting state at zero current; (2) the CIZRS in sufficiently large current; and (3) the nonreciprocal behavior. Moreover, the \textit{reciprocal} CIZRS and  hysteresis in current-resistivity measurements are predicted.
We also point out that the long-sought-after triangular finite-momentum superconducting state~\cite{Matsuda2007-em,Shimahara1998-iw}, whose order parameter is a {linear combination} of the three FF states, can be realized in valley-polarized superconductors.

\textit{The SLS model}---We begin with introducing a Hamiltonian which captures the symmetry and essential features of the twisted trilayer graphene/WSe$_2$ heterostructure. 
The Hamiltonian is 
$\hat{H}=\sum_{\bm{k}}\bm{c}^\dagger_{\bm{k}}H_{\bm{k}}\bm{c}_{\bm{k}}
+\hat{H}_{\rm int},$
where $\bm{c}^\dagger_{\bm{k}}=(c^\dagger_{\bm{k}\uparrow+},c^\dagger_{\bm{k}\downarrow+},c^\dagger_{\bm{k}\uparrow-},c^\dagger_{\bm{k}\downarrow-})$ describes the electrons with spin $s=\uparrow,\downarrow$ and valley $\eta=\pm$.
The normal-state Bloch Hamiltonian $H_{\bm{k}}$ is given by
\begin{equation}
H_{\bm{k}}=\begin{pmatrix}
    \varepsilon_{+,\bm{k}}&0\\
    0&\varepsilon_{-,\bm{k}}
\end{pmatrix}s_0,
\quad \varepsilon_{\eta,\bm{k}}=\xi_{\bm{k}}-\frac{\eta}{2}\delta\mu,
\end{equation}
and 
$\xi_{\bm{k}}=-t\sum_{n=0}^2\cos\left(\bm{k}\cdot C_3^n\hat{x}-\phi\right)-\mu.$
Here, $s_0$ represents the identity matrix in the spin space, while $C_3$ represents the anticlockwise threefold rotational matrix.
Thus, $C_3^n\hat{x}$ ($n=0,1,2$) with $\hat{x}=(1,0)$ are unit vectors pointing toward the three vertices of a regular triangle, and
the system accordingly has the $C_3$ symmetry.
The pairing interaction is given by
$\hat{H}_{\rm int}=-\frac{u}{V}\sum_{\bm{k}\bm{k}'\bm{q}\atop ss'\eta\eta'}c^\dagger_{\bm{k},s\eta}c^\dagger_{-\bm{k}+\bm{q},s'\eta'}c_{-\bm{k}'+\bm{q},s'\eta' }c_{\bm{k}',s\eta},
$
where $V$ denotes the system area and {$u$ denotes the on-site attractive interaction strength.} 
In the following,
we focus only on the intervalley spin-singlet $s$-wave channel, because pairing symmetries are expected to make only quantitative changes.

\begin{figure}
    \centering
\includegraphics[width=\linewidth]{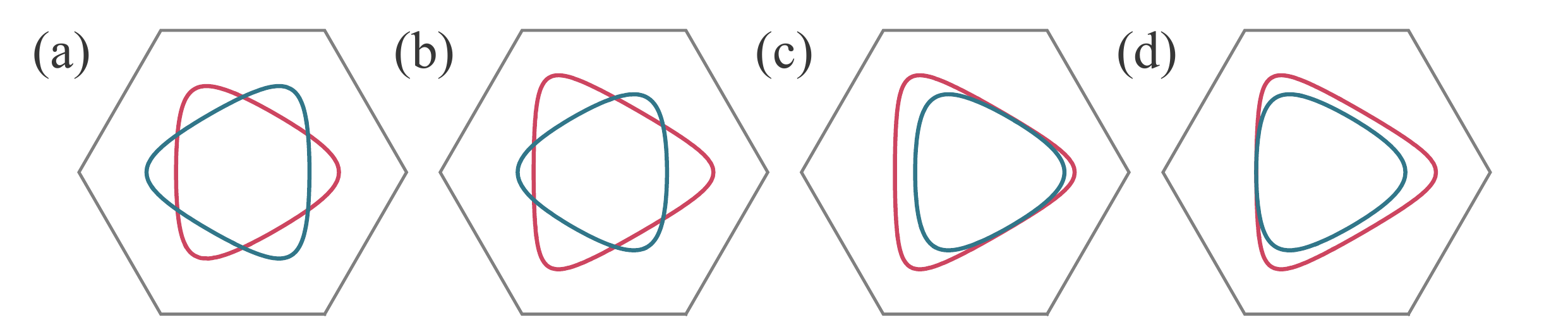}
    \caption{
    (a),(b) Fermi surfaces of the SLS model for (a) $\delta\mu=0$ and (b) $\delta\mu=0.35$. 
    Here and hereafter, we adopt $t=1$, $\mu=-0.65$, and $\phi=-0.2\pi$.
    The red and blue lines indicate Fermi surfaces of the valley $\eta=+$ and $-$, i.e., $\varepsilon_{+,\bm{k}}=0$ and $\varepsilon_{-,\bm{k}}=0$, respectively.
    (c),(d) Nestings of the Fermi surfaces for $\delta\mu=0.35$.
    Red and blue curves indicate $\varepsilon_{+,\bm{k}}=0$ and $\varepsilon_{-,-\bm{k}+\bm{q}}=0$, respectively, with (c) $\bm{q}=(0.33,0)$ and (d) $\bm{q}=(-0.19,0)$.
    }
    \label{fig:FSs}
\end{figure}

This model,
with some changes of notations, has been introduced by Scammell-Li-Scheurer (SLS) to discuss the intrinsic SDE in twisted trilayer graphene/WSe$_2$~\cite{Scammell2022-pv}.
Physically, 
$\eta=\pm$ specifies the graphene's valley degree of freedom, and $\bm{k}$ describes the wave number in the moir\'e Brilloiun zone.
Accordingly, $\delta\mu$ {characterizes the valley polarization and determines the difference in the chemical potential between} the two valleys ~\cite{Ajayan2016-cm, Novoselov2016-ev, Geim2013-vc, Novoselov2005-nv, Novoselov2004-pn}.
The valley polarization breaks both inversion and time-reversal symmetries, and is proposed as the origin of the observed SDE~\cite{Lin2022-cz,Scammell2022-pv,Banerjee2024-kd}  and Josephson diode effect in gate-defined Josephson junctions in twisted bilayer graphene~\cite{Diez-Merida2023-cw,Hu2023-cl}. 
The parameter $\phi$ {induces} the trigonal warping effect ~\cite{Scammell2022-pv}.
The {trigonal warping term causes} a threefold anisotropy as
shown in Figs.~\ref{fig:FSs}(a) and (b).
Thus, the SLS model captures the essential features of twisted trilayer graphene/WSe$_2$, despite its simplicity.

\textit{Finite-momentum superconductivity}---{We show in the following} that finite-momentum superconductivity can be realized in the SLS model. 
For this purpose, we introduce the pairing susceptibility
\begin{align}
\chi(\bm{q})=\int\frac{d^2k}{2\pi^2}\frac{1-f(\varepsilon_{+,\bm{k}})-f(\varepsilon_{-,-\bm{k}+\bm{q}})}{\varepsilon_{+,\bm{k}}+\varepsilon_{-,-\bm{k}+\bm{q}}},\label{eq:chiq}
\end{align}
where 
$f(\varepsilon)\equiv(e^{\varepsilon/T}+1)^{-1}$ represents the Fermi distribution at temperature $T$. The second-order superconducting transition takes place when $\chi(\bm{q})$ exceeds the inverse of the pairing interaction $1/u$ for a certain Cooper-pair momentum $\bm{q}$.
Finite-momentum superconductivity realizes if this happens at $\bm{q}\neq0$.

\begin{figure}
    \centering
\includegraphics[width=0.9\linewidth]{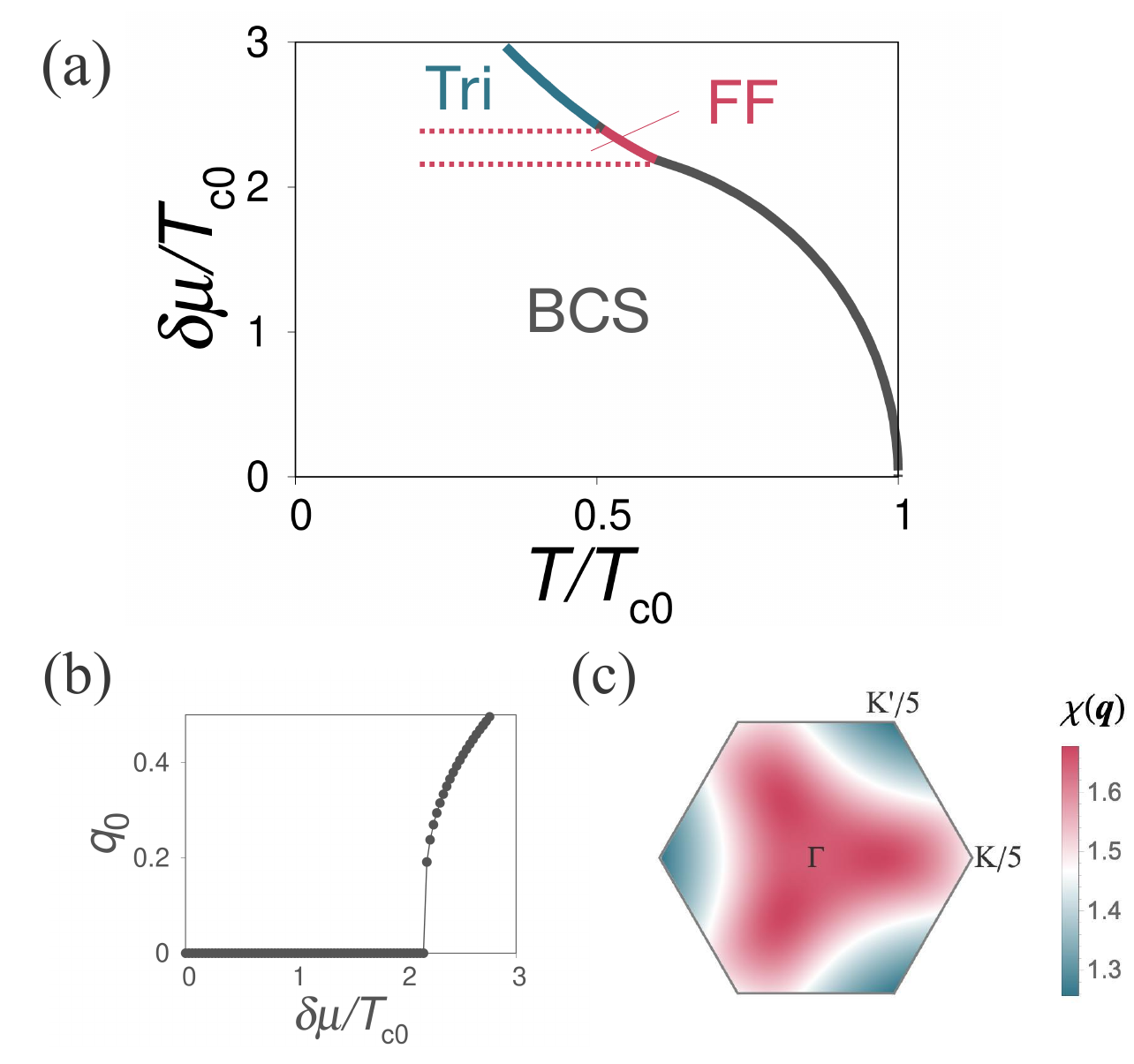}
    \caption{(a) The $T-\delta\mu$ phase diagram of the SLS model for $t=1$, $\mu=-0.65$, and $\phi=-0.2\pi$.
    We adopt $u=0.596$, %$u=1.192$, 
    which gives the transtion temperature $T_{\rm c0}\simeq 0.15$ for $\delta\mu=0$.
    The black, red, and blue solid lines indicate the second-order phase transition from the normal state to Bardeen-Cooper-Schrieffer (BCS), Fulde-Ferrell (FF), and triangular (Tri) states, respectively.
    The red dotted lines are the guide for the eye.
    (b) Equilibrium Cooper-pair momentum  along the transition line shown in panel (a).
    (c) Pairing susceptibility $\chi(\bm{q})$ near $\bm{q}=0$
    for  $\delta\mu=0.35$ and $T=0.08\simeq T_{\rm c}(\delta\mu)$.
    }    \label{fig:phase_diagram}
\end{figure}

We show in Fig.~\ref{fig:phase_diagram}(a) the transition line $T=T_{\rm c}(\delta\mu)$ determined by numerically solving {the equation} $1/u-\max_{\bm{q}}\chi(\bm{q})=0$.
The system realizes finite-momentum superconductivity
when valley polarization $\delta\mu$ exceeds a critical value $\delta\mu_{\rm c}\sim 2.2T_{{\rm c}0}$ [Figs.~\ref{fig:phase_diagram}(b) and (c)].
According to Fig.~\ref{fig:phase_diagram}(c), $\chi(\bm{q})$ for $\delta\mu\ge\delta\mu_{\rm c}$ reaches its maximum at three momenta $\bm{q}=\bm{q}_{0,1,2}$, where
\begin{align}
\bm{q}_0=q_0\hat{x},\quad \bm{q}_n=C_3^n\bm{q}_0\ (n=0,1,2).\label{eq:q0s}
\end{align}
This result is consistent with Ref.~\cite{Scammell2022-pv}, reproducing the appearance of finite-momentum superconductivity.
We note that Fig.~\ref{fig:phase_diagram}(b) shows a jump in $q_0$ at $\delta\mu_{\rm c}$, implying a first-order phase transition from the Bardeen-Cooper-Schrieffer (BCS) to finite-momentum states~\cite{Supplemental}.

The obtained finite-momentum superconductivity can be understood similarly to FF-Larkin-Ovchinnikov (LO) superconductors with fourfold anisotropy~\cite{Shimahara1999-of}. Generally speaking, the pairing susceptibility $\chi(\bm{q})$ is maximized by the best nesting, which is achieved by $\bm{q}$ {being} parallel to $\hat{x}$ rather than {being} antiparallel, as {shown} in Figs.~\ref{fig:FSs}(c) and \ref{fig:FSs}(d). Therefore, {a positive} $q_0>0$ and its $C_3$ equivalents are favorable. The obtained critical valley polarization $\delta\mu_{\rm c}=O(T_{{\rm c}0})$ is reasonable, since the splitting of the Fermi surfaces $\sim\delta\mu$ should exceed the Fermi-surface ambiguity $\sim T$ coming from the Fermi distribution.

\textit{Fulde-Ferrell and triangular states}---{In the above},  we identified the instability to finite-momentum pairing with three favorable momenta $\bm{q}_{0,1,2}$ defined in Eq.~\eqref{eq:q0s}.
This means that when $T$ is slightly below the transition temperature $T_{\rm c}(\delta\mu)$, the order parameter is generally the {linear combination of the three pairing order parameters such that}
\begin{align}
\Delta_{\rm a}(\bm{r})=\sum_{n=0}^2\Delta_ne^{i\bm{q}_n\cdot\bm{r}}.\label{eq:ansatz}
\end{align}
To identify the most stable combination, we substituted the ansatz Eq.~\eqref{eq:ansatz} for the mean-field free-energy density $F[\Delta]$, to obtain~\cite{Larkin1964-en,Shimahara1998-iw,Supplemental}
\begin{align}
F[\Delta_{\rm a}]
&=\alpha_0\,(|\Delta_0|^2+|\Delta_1|^2+|\Delta_2|^2)\notag\\
&\quad+\frac{1}{2}\beta(|\Delta_0|^2+|\Delta_1|^2+|\Delta_2|^2)^2\\
&\quad+\beta'(|\Delta_0|^2|\Delta_1|^2+|\Delta_1|^2|\Delta_2|^2+|\Delta_2|^2|\Delta_0|^2)\notag.
\end{align}
The {form of $F[\Delta_{\rm a}]$} is {dictated} by $C_3$ symmetry as well as the translational symmetry of the normal state.
The first Ginzburg-Landau (GL) coefficient $\alpha_0\equiv1/u-\chi(\bm{q}_0)$ is a small negative number, since we are focusing on {temperatures} slightly below $T_{\rm c}(\delta\mu)$, where the ansatz is justified.
The other GL coefficients are given by $\beta=\beta_{00}$ and $\beta'=2\beta_{10}-\beta_{00}$, with
\begin{align}
\beta_{nm}&=T\sum_{\omega_n}\int\frac{d^2k}{8\pi^2}\tr[g_{\bm{k}+\bm{q}_n}\,\bar{g}_{\bm{k}}\,g_{\bm{k}+\bm{q}_m}\,\bar{g}_{\bm{k}}]\label{eq:beta_GL}.
\end{align}
Here, we defined the Green's functions $g_{\bm{k}}=(i\omega_n-H_{\bm{k}})^{-1}$ and $\bar{g}_{\bm{k}}=(i\omega_n+\Theta H_{-\bm{k}}\Theta^{-1})^{-1}$ with Matsubara frequency $\omega_n$ and the time-reversal operator $\Theta$.

The stable order parameter is determined by $\beta'$.
In its absence, $F[\Delta_{\rm a}]$ is degenerate for all possible {linear combinations} of $\Delta_n$ due to the SU(3) symmetry.
The degeneracy is lifted by $\beta'$, and the FF state{, whose order parameter is defined by}
\begin{align}
\Delta_{{\rm FF},n}(\bm{r})=\Delta_ne^{i\bm{q}_n\cdot\bm{r}},\quad (n=0,1,2)\label{eq:FF_state}
\end{align}
is stabilized for $\beta'>0$. {This is because Eq.~\eqref{eq:FF_state} makes the $\beta'$ term vanish}.
On the other hand, the triangular state~\cite{Matsuda2007-em,Shimahara1998-iw} {defined by the order parameter}
\begin{align}
\Delta_{\rm Tri}(\bm{r})=\sum_n\Delta_n e^{i\bm{q}_n\cdot\bm{r}},\quad |\Delta_0|=|\Delta_1|=|\Delta_2|\label{eq:Tri_state}
\end{align}
is stable for $\beta'<0$ by maximizing the benefit of {the} $\beta'$ {term}.
The other solution given by the {linear combination} of two plane waves turns out to always have a subleading free energy~\cite{Supplemental}.

\begin{figure}
    \centering
    \includegraphics[width=0.8\linewidth]{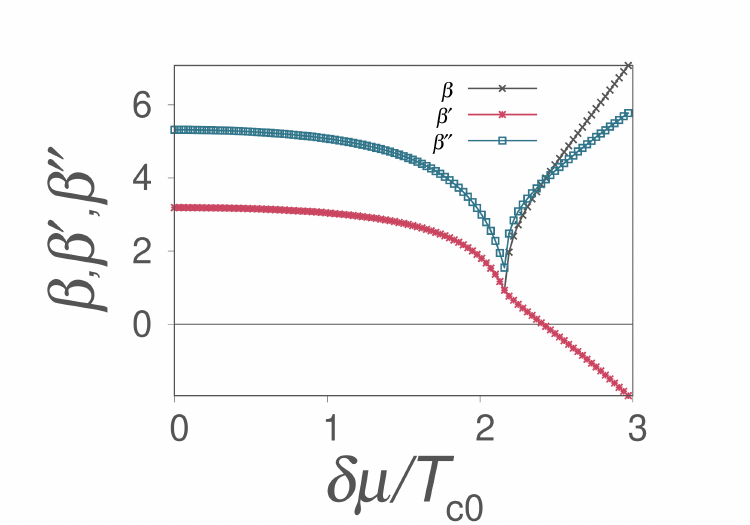}
    \caption{The quartic Ginzburg-Landau coefficients $\beta,$ (black) $\beta',$ (red) and $\beta''$ (blue) calculated along the transition line shown in Fig.~\ref{fig:phase_diagram}(a).
    }
    \label{fig:beta}
\end{figure}

We show in Fig.~\ref{fig:beta}
the numerical results of the GL coefficients $\beta$ and $\beta'$ of the SLS model along the transition line in Fig.~\ref{fig:phase_diagram}(a).
We also show 
$\beta''=\beta+2\beta'/3$,
whose positivity ensures the transition from the normal state to the triangular state to be second order~\cite{Supplemental}, while $\beta$ plays the same role for the FF state.
The black and red curves for $\beta$ and $\beta'$, respectively, coincide for small $\delta\mu$ since
$q_0=0$ and thus $\beta_{00}=\beta_{10}$ from Eq.~\eqref{eq:beta_GL}.
They split after $\delta\mu_{\rm c}$ is reached.
We can see that $\beta'$ stays positive {in the finite-momentum state} before
{the valley polarization reaches}
$\delta\mu\sim 2.4T_{\rm c0}>\delta\mu_{\rm c}$, while $\beta$ and $\beta''$ remain positive.
Thus, we obtain the phase diagram as shown in Fig.~\ref{fig:phase_diagram}(a), indicating the successive transitions of the BCS, FF, and the triangular states upon increasing the valley polarization $\delta\mu$ near the transition line.

The phase diagram can be understood based on the result of the isotropic 2D system in {a} Zeeman field. {In this case,} the system experiences successive transitions from the BCS, LO (indicating $\Delta(\bm{r})\sim\cos \bm{q}\cdot\bm{r}$~\cite{Larkin1964-en}), triangular, tetragonal, to the hexagonal state by  increasing the Zeeman field~\cite{Shimahara1998-iw,Matsuda2007-em}. By introducing the trigonal warping and breaking twofold rotational symmetry, {the} LO, tetragonal, and hexagonal states become unstable, while the triangular state should remain. Furthermore, the FF state may take the position of the LO state for small time-reversal breaking field. Indeed, {the} FF state is known to appear nearby the LO state in disordered $d$-wave superconductors~\cite{Agterberg2001-vo}. In this way,  the trigonal anisotropy and the valley polarization stabilize the FF and triangular states, giving exception to the common belief that the LO state is stable in most cases.

\textit{Current phase diagram of the FF states}---In the following, we focus on the parameter regime where {the} FF states are stabilized.
In this case, the system has the threefold degeneracy of the $\bm{q}_{0,1,2}$ states, and the actual state realized in experiments would be determined by perturbations of intrinsic and/or extrinsic origins.
Let us assume that small uniform nematicity $\epsilon$ is introduced to the system, e.g., by nematic order and/or uniaxial strain.
The free-energy density of each FF state changes from $F_0\equiv F[\Delta_{{\rm FF},n}]$ to
\begin{align}
F_n(\epsilon)=F_0+\epsilon\chi_n^{\rm nem},
\end{align}
with a nematic susceptibility $\chi_n^{\rm nem}$ of the $\bm{q}_n$ state.
This generally lifts the threefold degeneracy, and 
the state minimizing $F_n(\epsilon)$ is realized.

\begin{figure}
    \centering
\includegraphics[width=\linewidth]{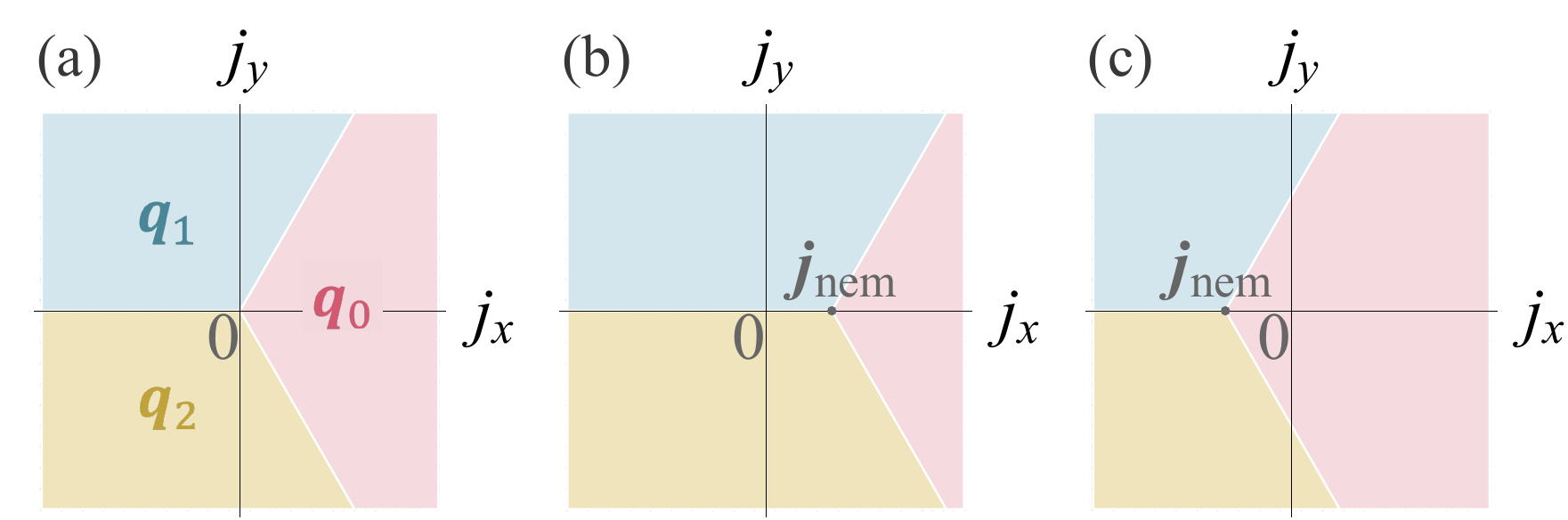}
    \caption{Thermodynamic phase diagrams of the spatially uniform FF states  in the current density $\bm{j}=(j_x,j_y)$ (a) without and (b),(c) with nematicity. 
    The panels (b) and (c) correspond to the cases of $\bm{j}_{\rm nem}$ parallel and antiparallel to $\hat{x}$.     The red, blue, and yellow regions indicate the FF states with Cooper-pair momentum $\bm{q}_0$, $\bm{q}_1$, and $\bm{q}_2$, respectively.
}
    \label{fig:pd_j}
\end{figure}

\begin{figure}
    \centering
\includegraphics[width=\linewidth]{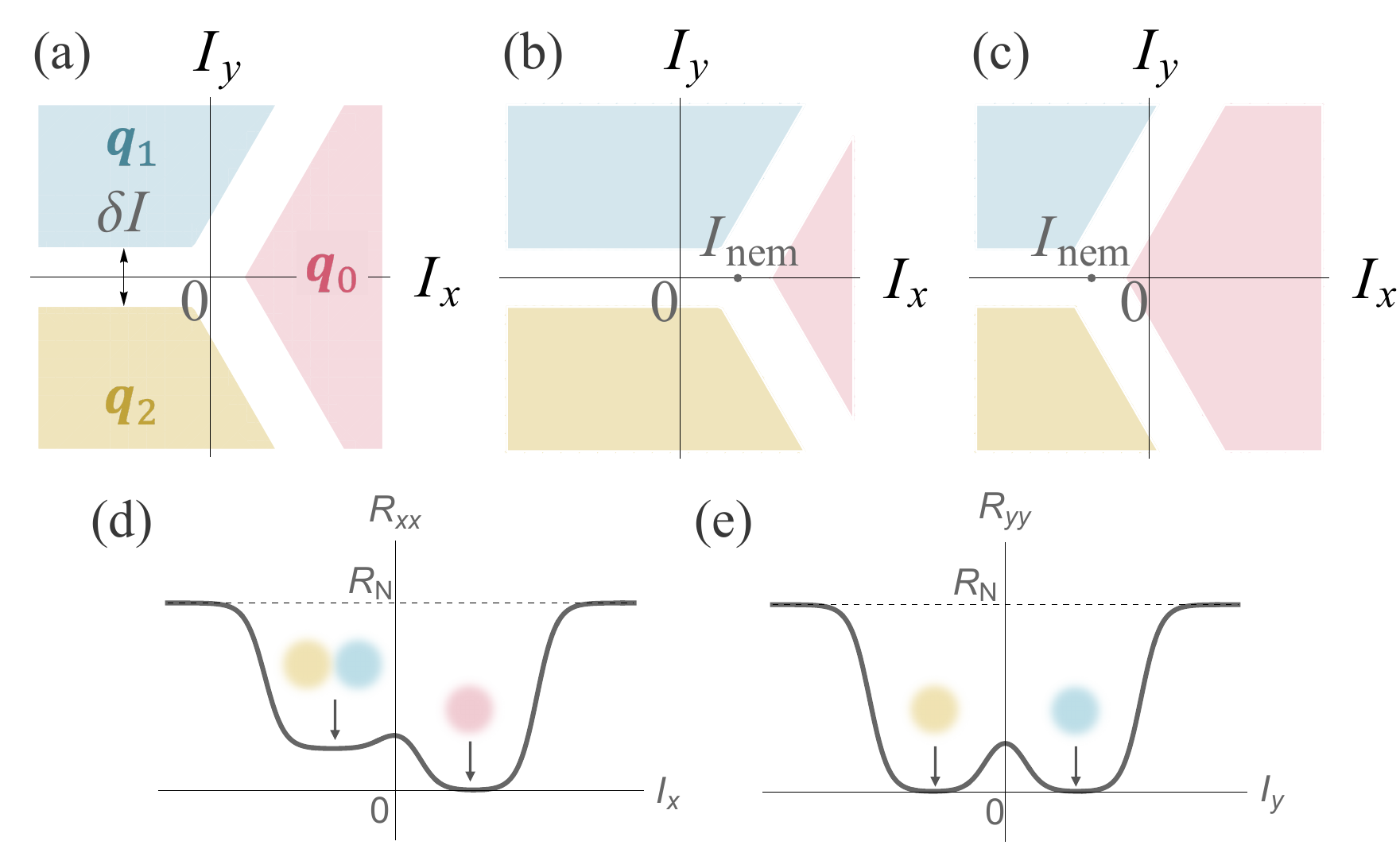}
    \caption{(a)-(c) Schematic $(I_x,I_y)$ phase diagrams and (d),(e) current-resistivity relation of the trigonal FF superconductors.
    Panels (a)-(c) correspond to Figs.~\ref{fig:pd_j} (a)-(c), and show the phase diagrams (a) with and (b),(c) without nematicity, with a phenomenologically introduced randomness $\delta I$.
    The white region indicates a multidomain state, and thus is resistive.
    Panels (d) and (e) indicate the expected resistance in $(I_x,0)$ and $(0,I_y)$ for panel (a), where $R_{\rm N}$ indicates the normal-state resistance.
    }
    \label{fig:pd_j_random}
\end{figure}

Interestingly, we can externally control the energetics of the FF states by using the supercurrent.
When the current density $\bm{j}$ is applied, the superconducting state is determined to minimize the Gibbs free-energy density~\cite{McCumber1968-ff, Samokhin2017-su, Tinkham2004-dh},
\begin{align}
G(\epsilon,\bm{j})&=\min_{n=0,1,2}[F_n(\epsilon)-\bm{q}_n\cdot\bm{j}+O(j^2,\epsilon j)].\label{eq:Gibbs}
\end{align}
Crucially, the momentum $\bm{q}_n$ of the FF state directly couples to the current density $\bm{j}$, and the competition with the nematicity triggers the switching between the FF states.
We show in Fig.~\ref{fig:pd_j}(a) the $(j_x,j_y)$ phase diagram in the absence of nematicity.
Each FF state is stabilized when the current is oriented along its Cooper-pair momentum.
By analyzing Eq.~\eqref{eq:Gibbs}, it turns out that the nematicity just shifts the entire phase diagram by a current density $\bm{j}_{\rm nem}=O(\epsilon)$, as discussed in  End Matter.
Figures~\ref{fig:pd_j}(b) and \ref{fig:pd_j}(c) illustrate the cases where
{$\bm{j}_{\rm nem}$ is parallel and antiparallel to the $x$ axis, with 
$\bm{j}_{\rm nem}=(j_{\rm nem},0)$ and $j_{\rm nem}=2(\chi_0^{\rm nem}-\chi_1^{\rm nem})\epsilon/3q_0$.
}
Note that we are focusing on the small-current regime, and the actual superconducting states would be {destroyed} at large current due to some critical-current mechanisms.
The thermodynamic $(j_x,j_y)$ phase diagram of the trigonal FF superconductors is one of the main results of this Letter.

\textit{Current-induced zero resistance state}---Keeping the bulk properties in mind, we point out that the switching phenomena of different FF states can naturally give rise to the CIZRS.
Let us assume that the twisted trilayer graphene/WSe$_2$ heterostructure stabilizes the threefold (near-)degenerate FF states as shown in the SLS model, and the domains of the FF states are formed after the temperature is lowered below a critical temperature.
This is a reasonable assumption for twisted trilayer graphene/WSe$_2$, because the valley polarization naturally follows from the field trainability of SDE and spin-orbit coupling~\cite{Lin2022-cz,Scammell2022-pv},  and  $\chi(\bm{q})$ with the three-peak structure has also been obtained in a  realistic continuum model 
for essentially the same reason as in the SLS model~\cite{Scammell2022-pv}.
Domains can be formed for various intrinsic and/or extrinsic reasons, including the configuration entropy and disorders in twist angles and/or local strain~\cite{Uri2020-aj,Lau2022-cs,Supplemental}.

Our scenario is based on two key observations.
First, the domains between FF states of different Cooper-pair momenta would be resistive.
Indeed, the Josephson energy $E_{\rm J}$ is tiny for such a situation~\cite{Yang2000-dr, Kaur2005-jf}.
The global coherence is not achieved when the typical value of $E_{\rm J}$ is smaller than temperature and/or the charging energy, according to the Josephson-network theory~\cite{Beloborodov2007-fm, Kapitulnik2019-tx}.
It is also known that vortices can be spontaneously formed at the interface of superconductors with different Cooper-pair momenta~\cite{Aoyama2012-sf}, whose motion may also produce dissipation. Thus, this multidomain state is expected to have a finite resistance lower than its normal-state value and realizes an incipient superconducting state with small resistance.

Second, the domain configuration would significantly be modified by large currents.
When we apply a current $I_x>0$ in the $x$ direction, the superconducting domains will switch to the $\bm{q}_0$ state according to,  e.g., Fig.~\ref{fig:pd_j}(a). The realized monodomain state would be dissipationless when zero resistance paths are formed. On the other hand, domains of $\bm{q}_1$ and $\bm{q}_2$ states would remain  if the current flows in the opposite direction with $I_x<0$, leaving the system resistive. Thus, nonreciprocal CIZRS is obtained in trigonal FF superconductors.

We expect that the actual switching occurs in a finite current strength, in contrast to Fig.~\ref{fig:pd_j}(a).
This is because there should be some domain pinnings as well as the variation in strength and direction of the current density each domain {experiences}.
The latter can occur by the current-path meandering, which is ubiquitous in percolating superconductivity and is particularly relevant to moir\'e superconductors~\cite{Yankowitz2019-wd,Uri2020-aj,Lau2022-cs,Balents2020-ke}.
These effects would broaden the transition lines between the different FF states, and we expect the $(I_x,I_y)$ phase diagram as in Fig.~\ref{fig:pd_j_random}(a).
Here, the white region indicates multidomain states and thus is resistive, and its width $\delta I$ is phenomenologically introduced as a measure of randomness.
The current-resistivity relation for $I_x$ would look like Fig.~\ref{fig:pd_j_random}(d), whose detailed shape can depend on the domain details, qualitatively explaining the nonreciprocal CIZRS observed in the experiment~\cite{Lin2022-cz}. 
An interesting prediction is the reciprocal CIZRS as in Fig.~\ref{fig:pd_j_random}(e), which is expected when a current {flows} in the $y$ direction {due to} the switching to $\bm{q}_1$ and $\bm{q}_2$ states in $I_y>0$ and $I_y<0$, respectively. 

Note that bulk nematicity has been observed in twisted trilayer graphene/WSe$_2$~\cite{Zhang2024-ec, Lin2022-cz, Siriviboon2021-un}.
In its presence, the $(I_x,I_y)$ phase diagram would be shifted by $I_{\rm nem}$, that is, a current scale corresponding to $\bm{j}_{\rm nem}$, leading to the phase diagrams as in Figs.~\ref{fig:pd_j_random}(b) and \ref{fig:pd_j_random}(c).
Interestingly, a substantial SDE with a monodomain configuration is expected instead of the CIZRS for $I_{\rm nem}\lesssim{-} \delta I$ as in Fig.~\ref{fig:pd_j_random}(c).
This is consistent with the experiment~\cite{Lin2022-cz}, where both huge SDE and CIZRS have been observed depending on the moir\'e filling and displacement field.
Such a state naturally accompanies a significant anisotropy in the critical current, which also agrees with a recent experiment~\cite{Zhang2025-fh}.

So far, we have assumed that each domain obeys the thermodynamic phase diagram given in Fig.~\ref{fig:pd_j}. On the other hand, the transitions between different FF states are first order, and thus hysteresis can appear depending on how the current is swept in experiments. When the current is reduced from a large value, e.g., in the positive $x$ direction, the sample can remain in the $\bm{q}_0$ state beyond the thermodynamic phase boundary~\cite{Supplemental}. We expect that this is related to the observed vanishing nonreciprocity and CIZRS when the sample was in a large current before the resistivity is measured~\cite{Lin2022-cz}.
More direct observation of the hysteresis in the current-resistivity relation, as well as the reciprocal CIZRS as in Fig.~\ref{fig:pd_j_random}(e), will give strong supporting evidence of our scenario.

\textit{Discussion}---While we have concentrated on the FF states [Eq.~\eqref{eq:FF_state}], the long-sought-after triangular state [Eq.~\eqref{eq:Tri_state}]~\cite{Shimahara1998-iw,Matsuda2007-em} is predicted to exist nearby according to the free energy analysis. Thus, valley-polarized superconductors offer a promising field-free platform for its observation. Considering its closeness to the FF states in the phase {diagram}, it could be accessible in twisted trilayer graphene/WSe$_2$ by tuning parameters. Being a honeycomb lattice of vortices and antivortices~\cite{Agterberg2011-rz}, the triangular state would be observable, e.g., by scanning-tunneling microscopy and superconducting quantum interference devices. The triangular state does not show degeneracy except for the overall translation (see End Matter for details), and will not show nonreciprocal CIZRS. This can be used as another test bed of our scenario.

Our prediction can be generalized to the other FF superconductors with multiple degeneracy.
For example, heterostructures of an Ising superconductor~\cite{Xi2015-lf, Lu2015-uu,Saito2015-aj} and ferromagnet or others may provide alternative platforms of nonreciprocal CIZRS~\cite{Supplemental}. In contrast to nonreciprocal CIZRS, a reciprocal CIZRS can be realized in $C_2$-symmetric FF superconductors such as disordered $d$-wave superconductors~\cite{Agterberg2001-vo}. Thus, it is interesting to explore this possibility in the candidate materials including CeCoIn${}_5$ thin films and superlattices~\cite{Naritsuka2021-ym}.

%%%%%%%%%%%%%%%%%%%%%%%%%%%%%%%%%

%\begin{acknowledgments}
\textit{Acknowledgments}---We appreciate fruitful discussions with J. I. A. Li. A.D. thanks Heng Wu and Shuntaro Sumita for helpful discussions, and Kazumasa Hattori and Satoru Hayami for informing him of Ref.~\cite{Agterberg2011-rz} and Refs.~\cite{Hayami2020-ix, Hayami2020-mu}. This work was supported by JSPS KAKENHI (Grant No. JP18H01178, No. JP18H05227, No. JP20H05159, No. JP21K18145, No. JP22H01181, No. JP22H04933,  No. JP21K13880, No. JP23K17353, No. JP24H01662, No. JP23KK0248, No. JP25H01249,  No. JP24H00007, No. JP24K21530). K. T. L. acknowledges the support of the Ministry of Science and Technology, China, and Hong Kong Research Grant Council through Grants No. MOST23SC01-A, No. 2020YFA0309600, No. RFS2021-6S03, No. C6025- 19G, No. C6053-23G, No. AoE/P-701/20, No. 16310520, No. 16307622, No. 16309223, and No.16311424.
%\end{acknowledgments}

%\bibliography{ref,ref_for_addition}

%

\appendix
\begin{center}\textbf{End Matter}\end{center}

\textit{Current phase diagram of the FF states}---Assuming the Fulde-Ferrell order parameter, the Ginzburg-Landau (GL) free-energy density can be written as
\begin{align}
F(\psi,\bm{q})\equiv F[\psi e^{i\bm{q}\cdot\hat{\bm{r}}}]=\alpha(\bm{q})|\psi|^2+\frac{\beta(\bm{q})}{2}|\psi|^4, \end{align}
where $\alpha(\bm{q})=1/u-\chi(\bm{q})$ and $\beta(\bm{q})$ is obtained by replacing $\bm{q}_0$ with $\bm{q}$ in $\beta_{00}$ defined in Eq.~\eqref{eq:beta_GL}.
Here, $F[\Delta]$ is the mean-field free-energy density functional measured from that of the normal state~\cite{Supplemental}.
At the transition temperature, $\alpha(\bm{q})$ changes its sign at momenta $\bm{q}_n\neq0$ with $n=0,1,2$.
After optimizing the order parameter amplitude $\psi=\psi(\bm{q})$, the GL free-energy density is given by
\begin{align}
F(\bm{q})&\equiv F(\psi(\bm{q}),\bm{q})\notag\\
&=\begin{cases}-{\alpha(\bm{q})^2}/{2\beta(\bm{q})}&(\alpha(\bm{q})<0)\\
0&(\alpha(\bm{q})\ge0)
\end{cases}.
\end{align}
When $T_{\rm c}(\delta\mu)-T\ll T_{\rm c}(\delta\mu)$, $\alpha(\bm{q})$ is negative only near $\bm{q}_n$.
Thus, we can concentrate on the three pockets in the $\bm{q}$ space around $\bm{q}_{0,1,2}$, where the free-energy density is expanded as
\begin{align}
F(\bm{q}_n+\delta\bm{q})=F_n+\frac{1}{2}D_n^{ij}\delta q_i\delta q_j+O(\delta q^3).
\end{align}
Here, $F_n=F(\bm{q}_n)=F[\Delta_{{\rm FF},n}]$ is the free energy of the $\bm{q}_n$ FF state, while
$D_n^{ij}\equiv \partial_{q_i}\partial_{q_j}F(\bm{q}_n)$ is its superfluid weight.

The Gibbs free energy is given by~\cite{McCumber1968-ff, Samokhin2017-su, Tinkham2004-dh}
\begin{align}
G(\bm{j})&=\min_{\bm{q}}[F(\bm{q})-\bm{q}\cdot\bm{j}]\notag\\
&\simeq\min_{n=0,1,2}\min_{\delta\bm{q}}[F_n+D_n^{ij}\delta q_i\delta q_j/2-(\bm{q}_n+\delta\bm{q})\cdot\bm{j}]\notag\\
&=\min_{n=0,1,2}[F_n-\bm{q}_n\cdot\bm{j}-\bm{j}^TD_n^{-1}\bm{j}/2].
\end{align}
In the presence of nematicity, we obtain
\begin{align}
G(\epsilon,\bm{j})=\min_{n=0,1,2}[F_0+\chi_n^{\rm nem}\epsilon-\bm{q}_n\cdot\bm{j}+O(j^2,\epsilon j)],
\end{align}
by replacing $F_n\to F_n(\epsilon)=F_0+\chi_n^{\rm nem}\epsilon$, where the $O(\epsilon j)$ contribution comes from the possible change of $\bm{q}_n$ by $O(\epsilon)$.

The minimization can be performed as follows.
We here consider a general nematicity and thus $\chi_{0,1,2}^{\rm nem}$ may take different values.
By using $\omega=e^{2\pi i/3}$, let us introduce
\begin{subequations}\begin{gather}
%\bar{\chi}\equiv\frac{\chi_0^{\rm nem}+\omega\chi_1^{\rm nem}+\omega^2\chi_2^{\rm nem}}{3},
\braket{\chi}\equiv(\chi_0^{\rm nem}+\chi_1^{\rm nem}+\chi_2^{\rm nem})/3,\\
\bar{\chi}\equiv(\chi_0^{\rm nem}+\omega\chi_1^{\rm nem}+\omega^2\chi_2^{\rm nem})/3,
\end{gather}\end{subequations}
%$\bar{q}_n\equiv \omega^n\bar{q}_0$, $\bar{q}_0=q_{0x}+iq_{0y}$,
and
\begin{align}
\bar{q}_n\equiv q_{nx}+iq_{ny}=\bar{q}_0\omega^n,\quad 
\bar{j}\equiv j_x+ij_y.
\end{align}
Then, we can write
\begin{align}
\chi_n^{\rm nem}=\braket{\chi}+2\Re[\omega^n\bar{\chi}^*],
\quad \bm{q}_n\cdot\bm{j}=\Re[\bar{q}_n\bar{j}^*],
\end{align}
by using  $1+\omega+\omega^2=0$,
and obtain
\begin{align}
% F_n+\chi_n^{\rm nem}\epsilon-\bm{q}_n\cdot\bm{j}&=F_0+2\Re[\omega^n(\bar{\chi}-q_0\bar{j})^*],\\
G(\epsilon,\bm{j})&=\min_{n=0,1,2}[F_0+\braket{\chi}\epsilon+\Re[\omega^n(2\bar{\chi}\epsilon-\bar{q}_0\bar{j}^*)]]\notag\\
&=\min_{n=0,1,2}[F_0+\braket{\chi}\epsilon-\bm{q}_n\cdot(\bm{j}-\bm{j}_{\rm nem})]\notag\\
&=G(0,\bm{j}-\bm{j}_{\rm nem})+\epsilon\braket{\chi}.
\end{align}
Thus, the minimization problem recasts to that of $\epsilon=0$, with the shift in the $(j_x,j_y)$ phase diagram by
\begin{align}
\bm{j}_{\rm nem}=2\epsilon(\Re[\bar{\chi}/\bar{q}_0^*],\Im[\bar{\chi}/\bar{q}_0^*]).
\end{align}
This conclusion applies to any trigonal FF superconductors beyond the SLS model.

Note that the SLS model preserves the $y$-mirror plane $M_y$ due to its simpleness.
When the nematicity $\epsilon$ preserves the $M_y$ symmetry, which corresponds, e.g., to the uniaxial strain in the $x$ direction, we obtain  $\chi_1^{\rm nem}=\chi_2^{\rm nem}$ and $\bar{\chi}=(\chi_0^{\rm nem}-\chi_1^{\rm nem})/3$, to reproduce $\bm{j}_{\rm nem}$ in the main text by using $\bar{q}_0=q_0$.
The minimization of $G(0,\bm{j})$ is then performed by
\begin{align}
G(0,\bm{j})&=F_0+\min_{n=0,1,2}\Re[-\bar{q}_0\omega^n\bar{j}^*]\notag\\
&=F_0+\min_{n=0,1,2}\left[-{q}_0j\cos\left(\frac{2\pi n}{3}-\theta\right)\right],
\end{align}
with $\bar{j}\equiv je^{i\theta}$ and $j\ge0$.
This clearly indicates the phase diagram as in the main text when $q_0>0$.

\textit{Degeneracy of the triangular state}---We briefly discuss the degeneracy of the triangular state.
Generally, the triangular state can be parametrized as
\begin{align}
\Delta_{\rm Tri}(\bm{r})&=|\Delta_0|\sum_{n=0,1,2}e^{i\bm{q}_n\cdot\bm{r}+i\varphi_n}.
\end{align}
Let us introduce 
%$\omega=e^{2\pi i/3}$ and $1+\omega+\omega^2=0$, as well as
$\braket{\varphi}=(\varphi_0+\varphi_1+\varphi_2)/3$ and
%\begin{align}
$\bar{\varphi}\equiv({\varphi_0+\omega\varphi_1+\omega^2\varphi_2})/{3}.$
%\end{align}
%We can also write
%\begin{align}
%\bar{q}_n\equiv q_{nx}+iq_{ny}=\omega^n\bar{q}_0.
%\end{align}
Then, we obtain
\begin{align}
\varphi_n&=\braket{\varphi}+2\Re[\omega^n\bar{\varphi}^*]\notag\\
&=\braket{\varphi}+\Re[\bar{q}_n(2\bar{\varphi}/\bar{q}_0^*)^*]\notag\\
&=\braket{\varphi}-\bm{q}_n\cdot\bm{r}_\varphi,
\end{align}
with 
\begin{align}
\bm{r}_\varphi\equiv-(\Re[2\bar{\varphi}/\bar{q}_0^*],\Im[2\bar{\varphi}/\bar{q}_0^*]).
\end{align}
Therefore, an arbitrary triangular state is written in the form
\begin{align}
\Delta_{\rm Tri}(\bm{r})=|\Delta_0|e^{i\braket{\varphi}}\sum_{n=0,1,2}e^{i\bm{q}_n\cdot(\bm{r}-\bm{r}_\varphi)}.
\end{align}
This means that the triangular state is not degenerate except for the overall phase and the choice of the origin, similarly to the LO state in systems with twofold anisotropy.

%%%%%%%%%%%%%%%%%%%%%%%%%%%%%%%%%%
%\input{Supplemental}
\begin{center}\textbf{Supplemental Material}\end{center}

\subsection{Derivation of the free-energy density}
We start from the mean-field Hamiltonian
\begin{align}
\hat{H}_{\rm mf}=\frac{1}{2}\bm{\Psi}^\dagger H_{\rm BdG}\bm{\Psi}+E_0.
\end{align}
Here, we defined the Nambu spinor $\bm{\Psi}^\dagger=(\bm{c}^\dagger,\bm{c}^TU_\Theta)$, the creation operator of electrons at position $\bm{R}$ and internal degrees of freedom  (spin and valley) $a$ by $[\bm{c}^\dagger]_{\bm{R},a}=c^\dagger_a(\bm{R})$, and the unitary part of the time-reversal operator $U_\Theta=is_y\eta_x$, where
$s_i$ and $\eta_i$ are the Pauli matrices in the spin and valley spaces.
The Bogoliubov-de-Gennes (BdG) Hamiltonian is given by 
\begin{align}
H_{\rm BdG}=\begin{pmatrix}
H&\Delta\\\Delta^\dagger&-\Theta H\Theta^{-1}
\end{pmatrix},
\end{align}
where $\Theta=U_\Theta K$ and $K$ represents the complex conjugation operator.
The constant $E_0$ for the SLS model is given by
\begin{align}
E_0=\frac{1}{4u}\Tr[\Delta^\dagger\Delta]+\frac{1}{2}\Tr[H],
\end{align}
where $\Tr[\cdot]$ runs over all the degrees of freedom including internal ones and the real-space coordinates.
The order parameter has the form
\begin{align}
\Delta=\sum_{\bm{q}}e^{i\bm{q}\cdot\hat{\bm{r}}}\Delta_{\bm{q}}.
\end{align}
Here, $\Delta_{\bm{q}}$ is just a scalar since the spin-singlet inter-valley s-wave order parameter is considered.
The position operator $\hat{\bm{r}}$ acts like $[\bm{c}^\dagger e^{i\bm{q}\cdot\hat{\bm{r}}}]_{\bm{R},a}=c^\dagger_a(\bm{R})e^{i\bm{q}\cdot(\bm{R}+\bm{r}_a)}$, where $\bm{R}+\bm{r}_a$ is the real-space position of the $(\bm{R},a)$ electron.
We have $\bm{r}_a=0$ for the SLS model.

The mean-field free energy is given by
\begin{align}
\Omega[\Delta]=E_0-\frac{T}{2}\sum_{\omega_n}e^{i\omega_n(+0)}\Tr'\ln G^{-1}(i\omega_n),
\end{align}
with $G^{-1}(i\omega_n)=i\omega_n-H_{\rm BdG}$.
Here, $\Tr'$ represents the trace over the Nambu degree of freedom as well as those of the normal state.
In the following, we abbreveate the argument $i\omega_n$ when it is unnecessary. 
Let us define
\begin{align}
\hat{G}_0\equiv \begin{pmatrix}
    g&0\\
    0&\bar{g}
\end{pmatrix},\quad \hat{\Delta}=\begin{pmatrix}0&\Delta\\
\Delta^\dagger&0\end{pmatrix},
\end{align}
by using the electron and hole Green's function $g=(i\omega_n-H)^{-1}$ and $\bar{g}=(i\omega_n+\Theta H\Theta^{-1})^{-1}$.
According to the Dyson equation $G^{-1}=\hat{G}_0^{-1}(1-\hat{G}_0\hat{\Delta})$, we obtain the Ginzburg-Landau expansion
\begin{align}
&\Omega[\Delta]-\Omega[0]\notag\\
&=\frac{1}{4u}\Tr[\Delta^\dagger\Delta]-\frac{T}{2}\sum_{\omega_n}e^{i\omega_n(+0)}\Tr'\ln(1-\hat{G}_0\hat{\Delta})\notag\\
&=\frac{1}{4u}\Tr[\Delta^\dagger\Delta]+\frac{T}{4}\sum_{\omega_n}\Tr'[(\hat{G}_0\hat{\Delta})^2]\notag\\
&\qquad+\frac{T}{8}\sum_{\omega_n}\Tr'[(\hat{G}_0\hat{\Delta})^4]+O(\Delta^6)\notag\\
&\equiv V(F_2[\Delta]+F_4[\Delta])+O(\Delta^6).
\end{align}
We defined the $O(\Delta^2)$ and $O(\Delta^4)$ terms as $F_2[\Delta]$ and $F_4[\Delta]$, respectively, with the system area $V$.
The superconducting contribution to the total free energy density is then given by
\begin{align}
F[\Delta]&=F_2[\Delta]+F_4[\Delta],
\end{align}
neglecting the $O(\Delta^6)$ contribution.
This is sufficient for our purpose since we are interested in the second-order superconducting transition.
By taking the trace over the Nambu degree's of freedom, the traces in $F_2[\Delta]$ and $F_4[\Delta]$ are evaluated as follows:
\begin{align}
\frac{T}{4}\sum_{\omega_n}\Tr'[(\hat{G}_0\hat{\Delta})^2]&=\frac{T}{2}\sum_{\omega_n}\Tr[g\Delta\bar{g}\Delta^\dagger],
\end{align}
and
\begin{align}
\frac{T}{8}\sum_{\omega_n}\Tr'[(\hat{G}_0\hat{\Delta})^4]&=\frac{T}{4}\sum_{\omega_n}\Tr[(g\Delta\bar{g}\Delta^\dagger)^2].
\end{align}

\subsection{$O(\Delta^2)$ term and finite-momentum superconductivity}
We first discuss the $F_2[\Delta]$ contribution.
The second-order superconducting transition occurs when this contribution turns negative for some $\Delta$.
Assuming the translational symmetry of the normal state, we  obtain
\begin{align}
F_2[\Delta]&=\sum_{\bm{q}}\frac{|\Delta_{\bm{q}}|^2}{V}\left[\frac{1}{4u}\Tr_{\rm N}1+\frac{T}{2}\sum_{\omega_n}\Tr[ge^{i\bm{q}\cdot\hat{\bm{r}}}\bar{g}e^{-i\bm{q}\cdot\hat{\bm{r}}}]\right]\notag\\
&\equiv \sum_{\bm{q}}\alpha(\bm{q})|\Delta_{\bm{q}}|^2.
\end{align}
Here, $\alpha(\bm{q})$ is given by $\alpha(\bm{q})=1/u-\chi(\bm{q})$, with
\begin{align}
\chi(\bm{q})&=-\frac{T}{2V}\sum_{\omega_n}\Tr[ge^{i\bm{q}\cdot\hat{\bm{r}}}\bar{g}e^{-i\bm{q}\cdot\hat{\bm{r}}}]\notag\\
&=-\frac{T}{2V}\sum_{\bm{k},\omega_n}\tr[g_{\bm{k}+\bm{q}}\bar{g}_{\bm{k}}].
\end{align}
Here, $\tr$ runs over the internal degrees of freedom, i.e., the spin and valley.
We obtain $\chi(\bm{q})$ in the main text by taking the trace and the summation over the Matsubara frequency.

As we have seen in the main text, the leading instability is achieved at three $\bm{q}$ vectors $\bm{q}_{n}=C_3^n\bm{q}_0$ ($n=0,1,2$) in the SLS model for large $\delta\mu$.
Cooper-pair momentum $q_0$ jumps at $\delta\mu=\delta\mu_{\rm c}$, which can be understood based on the GL theory with phenomenologically assuming the validity of gradient expansion.
According to the $C_3$ and $M_y$ symmetry, the GL coefficient $\alpha(\bm{q})$ takes the form
\begin{align}
\alpha(\bm{q})=\alpha_0+\frac{\alpha_2}{2}\bm{q}^2+\frac{2\alpha_3}{3}(q_x^3-3q_xq_y^2)+\frac{\alpha_4}{4}\bm{q}^4,
\end{align}
up to $O(q^4)$.
In contrast to the conventional FFLO states, the transition to the finite-momentum superconductivity takes place by the increase of $|\alpha_3|$ instead of sign reversal of $\alpha_2$, leading to the jump of $q_0$.
To see this, let us focucs on the $q_y=0$ line.
We obtain
\begin{align}
\alpha(q_x)=\frac{\alpha_2}{2}q_x^2+\frac{2\alpha_3}{3}q_x^3+\frac{1}{4}q_x^4.
\end{align}
Here, $\alpha_0$ is abbreviated since it does not affect the results.
We also assumed $\alpha_4>0$, and set it to unity by rescaling the momentum as well as $\alpha_2$ and $\alpha_3$.
The extrema are obtained by
\begin{align}
0=\alpha'(q_x)=q_x(\alpha_2+2\alpha_3q_x+q_x^2),
\end{align}
which has solutions other than $q_x=0$ at
\begin{align}
q_x=q_\pm\equiv -\alpha_3\pm\sqrt{\alpha_3^2-\alpha_2},
\end{align}
when $\alpha_3^2>\alpha_2$.
The value of $\alpha(q_x)$ for these solutions are
\begin{align}
\alpha(q_\pm)&=\frac{\alpha_3^2}{12}q_\pm^2(1\pm3\sgn[\alpha_3]D)(1\mp\sgn[\alpha_3]D),
\end{align}
with $D=\sqrt{1-\alpha_2/\alpha_3^2}$. Either of $\alpha(q_\pm)$ is smaller than $\alpha(0)=0$ when
\begin{align}
\alpha_3^2>\frac{9}{8}\alpha_2.
\end{align}
These results clearly indicate that first-order transition to $q_0\neq0$ occurs before $\alpha_2$ turns negative, since $\alpha_3=O(\delta\mu)$ and thus $\alpha_3^2>0$ in the presence of valley polarization.

\subsection{$O(\Delta^4)$ term and FF and triangular states}
When we assume the second-order phase transition from the normal state to the superconducting state, only plane waves with momenta near either one of $\bm{q}_n$ can appear in the solution of $\delta F[\Delta]/\Delta^\dagger=0$ with $F[\Delta]<0$.
As we approach the transition temperature from below, the ansatz
\begin{align}
\Delta_{\rm a}=\sum_{n=0,1,2}\Delta_ne^{i\bm{q}_n\cdot\hat{\bm{r}}},\quad \Delta_n\in\mathbb{C},
\end{align}
should be asymptotically correct.
The $O(\Delta^2)$ contribution is simply given by
\begin{align}
F_2[\Delta_{\rm a}]=\sum_n\alpha_0|\Delta_n|^2,
\end{align}
with $\alpha_0=\alpha(\bm{q}_n)$ owing to the $C_3$ symmetry.
We are focusing on such a temperature that $\alpha_0$ is a small negative number.

Below, we evaluate $F_4[\Delta_{\rm a}]$ to determine the stable order parameter.
Substituting the ansatz for $F_4[\Delta]$,
we obtain
\begin{align}
F_4[\Delta_{\rm a}]&=\frac{1}{2}\!\!\!\sum_{m_1,m_2,n_1,n_2\atop=0,1,2}\!\!\!\Delta_{m_1}^*\Delta_{m_2}^*\Delta_{n_1}\Delta_{n_2}\beta_{m_1m_2;n_1n_2},
\end{align}
with
\begin{align}
\beta_{m_1m_2;n_1n_2}&=\frac{T}{2V}\sum_{\omega_n}\Tr[ge^{i\bm{q}_{n_1}\cdot\hat{\bm{r}}}\bar{g}e^{-i\bm{q}_{m_1}\cdot\hat{\bm{r}}}\notag\\
&\qquad\cdot ge^{i\bm{q}_{n_2}\cdot\hat{\bm{r}}}\bar{g}e^{-i\bm{q}_{m_2}\cdot\hat{\bm{r}}}].
\end{align}
There are selection rules for $\beta_{m_1m_2;n_1n_2}$.
First, according to the translational symmetry of the normal state, only combinations of $m_1,m_2,n_1,n_2$ satisfying
\begin{align}
\bm{q}_{m_1}+\bm{q}_{m_2}=\bm{q}_{n_1}+\bm{q}_{n_2},
\end{align}
are allowed.
Note that $\bm{q}_n$ is not so large in our case that Umklapp processes do not occur.
It follows that
\begin{subequations}\begin{align}
\beta_a&\equiv\beta_{00;00}=\beta_{11;11}=\beta_{22;22},\\
\beta_b&\equiv\beta_{01;01}=\beta_{10;10}=\beta_{02;02}=\beta_{20;20}\notag\\
&=\beta_{12;12}=\beta_{21;21},\\
\beta_c&\equiv\beta_{10;01}=\beta_{01;10}=\beta_{20;02}=\beta_{02;20}\notag\\
&=\beta_{12;21}=\beta_{21;12},
\end{align}\end{subequations}
can be finite and the others vanish.
We used the relation
\begin{align}
\beta_{m_1m_2;n_1n_2}=\beta_{m_2m_1;n_2n_1},
\end{align}
which follows from
the cyclic property of the trace, and
the constraint from the $C_3$ symmetry
\begin{align}
\beta_{m_1m_2;n_1n_2}=\beta_{m_1+1,m_2+1;n_1+1,n_2+1},
\end{align}
where the subscripts are evaluated by modulo $3$.
Summing up all the contributions, we obtain
\begin{align}
F_4[\Delta_{\rm a}]&=\frac{\beta_a}{2}(|\Delta_0|^4+|\Delta_1|^4+|\Delta_2|^4)\\
&\ +(\beta_b+\beta_c)(|\Delta_0|^2|\Delta_1|^2+|\Delta_1|^2|\Delta_2|^2+|\Delta_2|^2|\Delta_0|^2).\notag
\end{align}
The GL coefficients are explicitly written down as
\begin{subequations}\begin{align}
\beta_a&=\frac{T}{2V}\sum_{\bm{k},\omega_n}
\tr[g_{\bm{k}+\bm{q}_0}\bar{g}_{\bm{k}}g_{\bm{k}+\bm{q}_0}\bar{g}_{\bm{k}}],\label{eq:beta_a}\\
\beta_b&=\frac{T}{2V}\sum_{\bm{k},\omega_n}
\tr[g_{\bm{k}}\bar{g}_{\bm{k}-\bm{q}_0}g_{\bm{k}}\bar{g}_{\bm{k}-\bm{q}_1}],\\
\beta_c&=\frac{T}{2V}\sum_{\bm{k},\omega_n}
\tr[g_{\bm{k}+\bm{q}_0}\bar{g}_{\bm{k}}g_{\bm{k}+\bm{q}_1}\bar{g}_{\bm{k}}].
\end{align}\end{subequations}
We can show $\beta_b=\beta_c$ by using the antiunitary property of $\Theta$.
Thus, we obtain
\begin{align}
F_4[\Delta_{\rm a}]
&=\frac{\beta}{2}(|\Delta_0|^2+|\Delta_1|^2+|\Delta_2|^2)^2\\
&\qquad+\beta'(|\Delta_0|^2|\Delta_1|^2+|\Delta_1|^2|\Delta_2|^2+|\Delta_2|^2|\Delta_0|^2),\notag
\end{align}
with
\begin{align}
\beta=\beta_a,\quad \beta'=\beta_b+\beta_c-\beta_a=2\beta_c-\beta_a.
\end{align}
This reproduces the expressions in the main text, by noting that $\beta_a=\beta_{00}$ and $\beta_c=\beta_{10}$.
{Note that we have not used the gradient expansion to obtain $F_4[\Delta_{\rm a}]$, which could be important because the gradient expansion may not be justified for  description of the first-order phase transition of the Cooper-pair momenta.}

Now the minimization of $F[\Delta_a]$ recasts to that by $\Delta_0$, $\Delta_1$, and $\Delta_2$.
It is convenient to parametrize them as
\begin{align}
&(\Delta_0,\Delta_1,\Delta_2)\notag\\
&=R(e^{i\varphi_1}\cos\theta,e^{i\varphi_2}\sin\theta\cos\phi,e^{i\varphi_3}\sin\theta\sin\phi).
\end{align}
We obtain
\begin{align}
F[\Delta_{\rm a}]
%&=\alpha_0R^2+\frac{\beta}{2}R^4+\beta'R^4(\cos^2\theta\sin^2\theta\cos^2\phi\\&\quad+\sin^4\theta\sin^2\phi\cos^2\phi+\cos^2\theta\sin^2\theta\sin^2\phi)\\
&=\alpha_0R^2+\frac{\beta}{2}R^4\notag\\
&\quad+\beta'R^4\left[\sin^2\theta+\sin^4\theta\left(\frac{1}{4}\sin^22\phi-1\right)\right].
\end{align}
From symmetry, we can focus on the exterma in the region $0\le\theta\le\pi/2$ and $0\le\phi\le\phi/2$.
They are given by (i) $\theta=0$, (ii) $\phi=0,\pi/2$ with $\theta=\pi/4$, and (iii) $\phi=\pi/4$ with $\sin^2\theta=2/3$.
The states characterized by the conditions (i), (ii), and (iii) correspond to FF, double-$q$, and triangular states, respectively.
Here, double-$q$ state (DQ) is defined as 
\begin{align}
\Delta_{{\rm DQ},n}(\bm{r})=\sum_{m\neq n}|\Delta_m|e^{i\bm{q}_m\cdot\bm{r}},\quad |\Delta_{n+1}|=|\Delta_{n-1}|.
\end{align}
The free-energy density corresponding to these states are obtained by optimizing
\begin{subequations}\begin{align}
F_{\rm FF}(R)&=\alpha_0R^2+\frac{\beta}{2}R^4,\\
F_{\rm DQ}(R)&=\alpha_0R^2+\frac{\beta+\beta'/2}{2}R^4,\\
F_{\rm Tri}(R)&=\alpha_0R^2+\frac{\beta+2\beta'/3}{2}R^4,
\end{align}\end{subequations}
in terms of $R$.
This expression indicates that $\beta>0$, $\beta+\beta'/2>0$, and $\beta''\equiv\beta+2\beta'/3>0$ are required for the transition from the normal state to FF, DQ, and Tri states to be second order, respectively.
After optimization by $R$, we conclude
\begin{subequations}\begin{gather}
F_{\rm FF}=-\frac{\alpha_0^2}{2\beta},\quad F_{\rm DQ}=-\frac{\alpha_0^2}{2(\beta+\beta'/2)},\\
F_{\rm Tri}=-\frac{\alpha_0^2}{2(\beta+2\beta'/3)},
\end{gather}\end{subequations}
and therefore
\begin{subequations}\begin{align}
&F_{\rm FF}<F_{\rm DQ}<F_{\rm Tri}\quad (\beta'>0),\\
&F_{\rm Tri}<F_{\rm DQ}<F_{\rm FF}\quad (\beta'<0).
\end{align}\end{subequations}

\section{Existence of FF domains and their evolution in external current}
{Here we provide remarks on the presence of FF domains.
It has been argued in Refs.~\cite{Izyumov2000-og,
Fominov2003-ju,
Izyumov2002-kc} that domains between the zero- and finite-momentum states cannot exist at a superconductor/ferromagnet (S/F) heterostructure.
This statement apparently contradicts with our assumption of the FF domains; however, the two physical systems are essentially different and thus the result does not apply to our system.
The key difference  is that in the S/F heterostructure, the ferromagnet is not superconducting and the pairing must be induced by the superconductor with zero-momentum pairing. 
In our case, the superconducting state itself has three degenerate solutions with three different pairing momenta. This difference naturally leads to different conclusions regarding the existence of domains with non-equivalent pairing momenta.}

{To give a physical picture about how the domains with different pairing momenta can be formed, let us assume that there is a weak nematic disorder in the region $x>0$ and $x<0$ and, for the time being, that the regions $x>0$ and $x<0$ are decoupled. As discussed in the main text, such different nematicity lifts the degeneracy and favors different FF states with different pairing momenta. For example, $\bm{q}_0$ and $\bm{q}_1$ states are realized for $x>0$ and $x<0$, respectively. Due to the decoupling of the two regions, this is the exact ground state of the total system. The possibility of FF domains can be tested by switching on the coupling of the two regions and seeing whether the ground state is significantly altered. Clearly, the free energy of each bulk-sized domain dominates over that from the interface, and the system remains to have $\bm{q}_0$ state for $x>0$ and $\bm{q}_1$ state for $x<0$ as long as $x$ is away from the boundary. Thus, existence of FF domains is ensured as long as each domain is sufficiently large. While the actual size of FF domains would depend on system details and is difficult to be determined, it is natural to assume that they are sufficiently large, considering that each superconducting island in moir\'e systems may be as large as 1$\mu$m~\cite{Uri2020-aj}. While we have illustrated the case with the domain pinning due to nematic disorder, domains can also appear due to entropy reasons even in clean systems, as is the case for the ferromagnetic domains. }

{We also comment on the dynamics of each domain in the presence of external current.
In contrast to the S/F heterostrucuture, whose dynamics could be very complicated, the evolution of the FF domains can be determined by thermodynamics.
Summarizing the discussions in the main text, in the presence of an applied current, the superconducting domains with $\bm{q}_n$ that minimizes the Gibbs free energy~\cite{McCumber1968-ff, Samokhin2017-su, Tinkham2004-dh} will grow in size, as indicated by the phase diagrams in Figs.4(a)-(c). 
}

\section{Supplementary explanation of the expected resistance hysteresis}
\begin{figure}
    \centering
\includegraphics[width=0.8\linewidth]{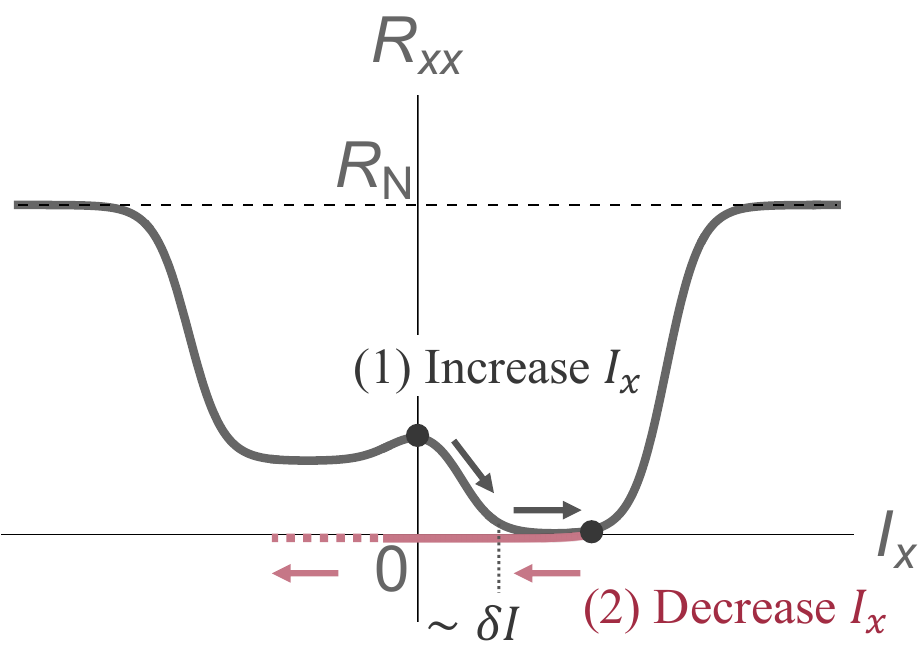}
    \caption{
Schematic figure for the expected hysteresis in the resistivity measurement. 
    }
    \label{fig:S1}
\end{figure}

Here, we make a supplementary explanation of the predicted hysteresis in the resistance measurement.
The situation is schematically shown in Fig.~\ref{fig:S1}. 
We start with $I_x\sim0$, wherein the system shows a finite resistance due to the presence of the domains with different FF states. We first increase the current $I_x$ to exceed $\sim\delta I$ [step (1)]. Then, the domains with the Cooper-pair momentum $\bm{q}_0$ will grow in size, and a zero-resistance state will appear, as indicated in the phase diagram Fig. 5(a) in the main text.
Next, we reduce the current [step (2)]. 
Even when $I_x$ is reduced below $\sim\delta I$, it is possible that the domains with the Cooper-pair momentum $\bm{q}_0$ continue to form a current path, and therefore zero-resistance state may appear even when $I_x$ is much smaller than $\delta I$.
Observation of this kind of hysteresis will support our scenario of the nonreciprocal CIZRS.

\section{Alternative platforms of nonreciprocal CIZRS}
As discussed in the main text, our prediction of CIZRS can be generalized to the other FF superconductors with multiple degeneracy.
For example, it is possible to obtain an equivalent to the SLS model by considering an Ising superconductor~\cite{Xi2015-lf, Lu2015-uu,Saito2015-aj}/ferromagnet heterostructure as is briefly discussed below.
In addition to the Ising superconductor/ferromagnet heterostructures,
$f$-wave altermagnets~\cite{Das2024-qn} and $120^\circ$ antiferromagnets~\cite{Hayami2020-ix, Hayami2020-mu} could offer platforms when sandwiched by a superconductor and a ferromagnet, since they show Fermi surfaces similar to Fig.~1(a)~\cite{Das2024-qn,Hayami2020-ix, Hayami2020-mu}.

In the following, we briefly show that an equivalent of the SLS model can be obtained based on the heterostructure of an Ising superconductor and a ferromagnet.
The Hamiltonian of such a system can be written as
\begin{align}
\hat{H}=\sum_{\bm{k}}\bm{c}^\dagger_{\bm{k}}H(\bm{k})\bm{c}_{\bm{k}}+\hat{H}_{\rm int},
\end{align}
with
\begin{align}
H(\bm{k})=E_{\bm{k}}+g_z(\bm{k})s_z-hs_z,
\end{align}
where $E_{\bm{k}}$, $g_z(\bm{k})$, and $h$ represent the hopping energy, Ising spin-orbit coupling, and the exchange field in the $z$ direction introduced by the proximity to the ferromagnet. 
Here, $\bm{k}$ is the usual wave number measured from the Gamma point, and
$\xi(\bm{k})$ and $g_z(\bm{k})$ are given by
\begin{align}
E(\bm{k})&=-t_\parallel\left(\cos k_x+2\cos\frac{\sqrt{3}k_y}{2}\cos\frac{k_x}{2}\right)-\mu\notag
\\&=-t_\parallel\sum_{n=0}^2\cos(\bm{k}\cdot C_3^n\hat{x})-\mu,\\
 g_z(\bm{k})&=\alpha_{\rm I}\left(\sin k_x-2\sin\frac{k_x}{2}\cos\frac{\sqrt{3}}{2}k_y\right)\notag
 \\&=\alpha_{\rm I} \sum_{n=0}^2\sin(\bm{k}\cdot C_3^n\hat{x}),
%(\sin k_x^3-3\sin k_y^2\sin k_x).
\end{align}
assuming the nearest-neighbor bonds of the triangular lattice.
By rewriting
\begin{align}
t_\parallel=t\cos\phi,\,\alpha_{\rm I}=-t\sin\phi,\, \text{and} \, h=-\delta\mu/2,
\end{align}
we obtain
\begin{align}
H(\bm{k})
%&=-t\sum_{n=0}^2\cos(\bm{k}\cdot C_3^n\hat{x})\cos\phi+\sin(\bm{k}\cdot C_3^n\hat{x})s_z\sin\phi-\mu-hs_z\\
%&=-t\sum_{n=0}^2\cos(\bm{k}\cdot C_3^n\hat{x}-\phi s_z)-\mu-hs_z\\
&=-t\sum_{n=0}^2\cos(s_z\bm{k}\cdot C_3^n\hat{x}-\phi)-\mu+\delta\mu s_z/2\notag\\
&=\begin{pmatrix}
\varepsilon_{+\bm{k}}&0\\
0&\varepsilon_{-\bm{k}}
\end{pmatrix},
\end{align}
and thus the SLS model is realized in the spin space instead of the valley space.
The pairing interaction is also equivalent by assuming the on-site attractive interaction and spin-singlet $s$-wave superconductivity, and thus the phase diagram is exactly the same as that of  the SLS model when the corresponding parameters are equivalent.

\end{document}